\documentstyle[preprint,revtex]{aps}
\begin{document}
\draft
\begin{title}
 Maximum Azimuthal Anisotropy of Neutrons from Nb-Nb Collisions \\
at 400 AMeV and the Nuclear Equation of State
\end{title}
\author{M. Elaasar$^{(1)}$, R. Madey$^{(1)}$, W. M. Zhang$^{(1)}$, 
J. Schambach\cite{joe}$^{(1)}$, D. Keane$^{(1)}$, \\
B. D. Anderson$^{(1)}$, 
A. R. Baldwin$^{(1)}$, J.W. Watson$^{(1)}$, G. D. Westfall$^{(2)}$, \\
G. Krebs$^{(3)}$, H. Wieman$^{(3)}$, C. Gale$^{(4)}$ and K. Haglin$^{(4)}$}

\begin{instit}
$^{(1)}$Kent State University, Kent, OH 44242-0001
\end{instit}

\begin{instit}
$^{(2)}$Michigan State University, East Lansing, MI 44824-1321
\end{instit}
\begin{instit}
$^{(3)}$Lawrence Berkeley Laboratory, Berkeley, CA 94720
\end{instit}
\begin{instit}
$^{(4)}$McGill University, Montr\'eal, Qu\'e., Canada H3A-2T8.
\end{instit}
\begin{abstract}
We measured the first azimuthal distributions of triple--differential 
cross sections of neutrons emitted in heavy-ion collisions, and  compared 
their maximum 
azimuthal anisotropy ratios with Boltzmann--Uehling--Uhlenbeck (BUU) 
calculations with a momentum-dependent interaction. 
 The BUU calculations agree with the triple- and 
double-differential cross sections for positive rapidity neutrons emitted at 
 polar angles from 7 to 27 degrees; however, the  
maximum azimuthal anisotropy ratio for these free neutrons is 
insensitive to the size of the nuclear incompressibility modulus K
characterizing the nuclear matter equation of state. 
\end{abstract}

	An important goal of relativistic heavy-ion physics is to extract
information on
the equation-of-state(EOS) of nuclear matter, which relates
density, temperature, and pressure. This relationship has drastic
implications for heavy-ion collisions and also has important
astrophysical consequences\cite{greiner}. As far as bulk properties are
concerned, the vast body of knowledge
associated with nuclear structure physics contributes information on
a single point in the
 temperature-density plane  [viz., at ~T~=~0 and ~$\rho$~=~$\rho_0$~], where
$\rho_0$ is the density of nuclear matter at equilibrium.
Heavy-ion collisions provide a laboratory tool for investigating
the behavior of nuclear matter in regions of temperature and
density removed from equilibrium. In recent years, progress
towards this goal has occurred in both theory and
experiment;  however, the task of deducing
the EOS from high-energy heavy-ion collisions is a difficult one. One
 complication arises because nonequilibrium aspects of the
collision are important\cite{gale90}; this fact
reflects on
certain observables that were thought originally to be good probes
of the EOS such as, for example, transverse momentum
generation\cite{gale90,gale89,doss86},
 pion production\cite{stock82}, kaon production\cite{sch87}, and flow-angle
distributions\cite{gusta84}. In this Letter, we concentrate on
 another observable associated with the azimuthal distribution about a
reaction plane for particles measured in a heavy-ion collision. Such azimuthal
distributions were measured \cite{previous}
and  calculated\cite{sch87} previously for charged particles; but here we
 pursue the sensitivity to the EOS of the maximum azimuthal anisotropy
ratio\cite{Welke88} (defined below) for neutrons.
We might expect the azimuthal distributions of neutrons to differ from 
those for charged particles primarily because of the absence of Coulomb 
 effects and different nucleon-nucleon cross sections. 
First, we address experimental observations, and then compare the data with
calculations based on nuclear transport theory.
 We find that the experiment is insensitive to the nuclear incompressibility
modulus K in the EOS.

The experiment  was designed to measure 
triple-differential cross-sections
 $d^3\sigma$/$d(cos\theta)$ $d(\phi- \phi_{R})$ $d\alpha$ for neutrons from
high-multiplicity
 collisions of equal-mass nuclei as a function of mass number and bombarding
energy. 
The symbol $\alpha$ \ $(\equiv Y/Y_P)_{CM}$ denotes the neutron 
 rapidity normalized to the projectile rapidity $Y_{P}$
in the center-of-mass (CM) system. We obtained an estimate of the azimuthal 
angle $\phi_{R}$
 of the reaction plane by measuring the transverse velocities
of charged particles emitted with positive rapidities above a cutoff rapidity
in the CM 
 system and summing them to obtain a total transverse-velocity vector.
 The $\phi$ -distribution of these summed transverse-velocity vectors is peaked
about the reaction plane. This transverse-velocity method\cite{Fai87} is an 
adaption of the transverse-momentum method of Danielewicz and 
Odyniec\cite{dan85}.
As described previously\cite{WM}, we determined $\phi_R$ with a dispersion
$\Delta$ $\phi_R$ ( $ \ge$ $\alpha_c$) for charged fragments above a normalized
rapidity  $\alpha_c$. For 400 AMeV Nb-Nb, we observed $\Delta$ $\phi_R$ 
$\sim 40^{\circ}$ for all $\alpha_c$ with uncertainties of about 
5 \%.

Projectile ions traversed a beam telescope before interacting with 
the target. An array of 18 scintillation detectors was 
used to detect neutrons emitted from each interaction.
The size of each neutron detector and the flight path at each angle were
selected to provide approximately equal counting rates and energy 
resolutions for the highest energy neutrons of interest at each angle.
 These mean-timed\cite{baldwin,Madey1} detectors spanned the polar angles
 from $3^\circ$ to $90^\circ$. The flight paths to
the 18 detectors ranged from 5.91 to 8.38 m.
Charged particles incident on each of the 18 neutron detectors were vetoed
with either a 6.3- or  9.5-mm thick anticoincidence plastic scintillator.
The time-of-flight (TOF) of each detected neutron was determined by measuring
the time difference between the detection of a neutron in one of the neutron
detectors and the detection of a Nb ion in a beam telescope. 
A {\it plastic-wall} array, 5-m wide and 4.5-m high, of 184 scintillation 
detectors (each 9.5-mm thick) was used to detect charged fragments emitted 
from each interaction. The multiplicity of the detected charged
fragments indicated the degree of centrality of the collision.
The flight paths to the 184 detectors ranged from 4.0 
to 5.0 m. The velocity of each charged fragment detected in the plastic wall
in each collision was extracted from the measured
flight-time. The uncertainty in the velocity was typically 3\% for  400  AMeV
charged fragments detected in the plastic wall;
about two thirds of the uncertainty came from the
intrinsic time dispersion of the detectors, and another one third from the
uncertainty in the position of the charged fragments.
A thin steel (0.95-mm) sheet, which covered the front side of the 24
inner detectors, was used to absorb the $\delta$-rays produced as the beam
traversed the air after it exited the beam pipe just upstream of the target.
The Nb target (with a physical thicknesses of 2.04 g/cm$^2$) was oriented
 at $60^{\circ}$ with respect to the beam. The beam energy at the center of the
target was 400 AMeV with an energy spread of $\pm 41$ AMeV.
Auxiliary measurements with steel shadow shields, typically 1.0-m long, were
used to determine target-correlated backgrounds. Each shadow shield was
 located  approximately halfway between the target and the detector.
The shadow shields attenuated neutrons by a factor $> 10^3$ at all energies.

To compare our experimental results with transport model 
calculations, it is necessary to match the impact 
parameter range in the model with the observed multiplicity values.
Well--known geometrical arguments\cite{impact} for estimating the impact 
parameter assume a correlation between the impact parameter and the
fragment multiplicity $M$. Measurements of the multiplicity distribution with
the target removed showed that collisions with a charged
multiplicity $M \ge $ 26 contained only about 5\% of
the background contamination from collisions of Nb with the air or the
material in the beam telescope.
This multiplicity threshold  selects about 22\% of the
total geometric cross-section which corresponds to a value of 0.47 for the
ratio of the maximum impact parameter $b_0$
to the nuclear diameter 2R.

 Welke et al.\cite{Welke88} discussed the maximum azimuthal anisotropy ratio
as a testing ground for the EOS. 
 At each polar angle, the maximum azimuthal anisotropy ratio $r(\theta,\alpha)$
 for neutrons in a given rapidity bin is the ratio of the {\it maximum} value
of the triple-differential cross section $\sigma_3^{MAX}$ to the 
{\it minimum} value $\sigma_3^{MIN}$.
For positive (negative) CM rapidities, we observed that
$\sigma_3$ reaches a maximum (minimum) at $\phi~ =~ 0^{\circ}$ ($180^{\circ}$) 
 and a minimum (maximum) at $180^{\circ}$ ($ 0^{\circ}$).
To evaluate  the ratio $r(\theta)$ for each rapidity bin, we fit the measured
$\sigma_3$ at each $\theta$ with a function of the form
$ \sigma_3(\phi,\theta) = a(\theta) + b'(\theta) \cos(\phi -\phi_R)$, and
obtain an estimate $r'(\theta)$.  For positive rapidities, the parameter 
 $b'(\theta)$ is positive; thus, $r'(\theta, \alpha > 0)$ = 
 [$a(\theta) + b'(\theta)$] /[$a(\theta) - b'(\theta)$].
 To correct for the finite rms dispersion in $\phi_R$, we use the fact that 
the parameter $b'(\theta)$ = $b(\theta) exp[-{1 \over 2}(\Delta \phi_R)^2]$
for a Gaussian distribution of $\phi_R$ with an rms dispersion $\Delta \phi_R$
. Finally, from the observed $b'(\theta)$ and the measured dispersion
 $\Delta \phi_R$, we calculate $b(\theta)$ and obtain the 
dispersion-corrected ratio
 $r(\theta)$ = [$a(\theta) + b(\theta)$]/ [$ a(\theta) - b(\theta)$].

For theoretical interpretation, we rely here on the BUU approach\cite{BUU} 
with a momentum--dependent nuclear mean field, 
$U (\rho ,{\vec p} )$, as parameterized in Ref \cite{gale90}. The 
momentum-dependent interaction 
  is essential not only 
from a theoretical standpoint\cite{jeukenne},
but also  has important observable implications\cite{momdep,pd93}. 
The BUU calculations here
have implemented a new algorithm, which considers the altered in--medium
phase space density that occurs with a momentum--dependent interaction.
This algorithm amounts to having a scattering cross--section that depends on
 the effective mass\cite{pand92} of the nucleon. This correction turns out to
be non--negligible in our case and a systematic investigation of its experimental
consequences will appear elsewhere\cite{hag}.
Transport theories used to extract the EOS must account for the polar-angle
dependence of the maximum azimuthal anisotropy ratio $r(\theta)$;
furthermore, at each polar angle the calculations should fit the absolute 
 triple-differential cross-sections 
 and the absolute double-differential cross sections (obtained by 
integrating the triple-differential cross-sections over the azimuthal angle).
It is important to point out that such complete
data provide a crucial test for theoretical models.

The full one--body BUU theory considers all nucleons emitted in a
nucleus--nucleus collision: free nucleons 
 and  those carried away in composite fragments. The initial application of the
full model revealed that the maximum neutron azimuthal anisotropy ratio 
is sensitive to the value of K, as in the original 
proposal\cite{Welke88}; however, we observed that these 
analyses were dominated by
 a trend not present in the
experimental data\cite{RM93} which include only {\it free}
neutrons. In an attempt to gain further 
understanding, we proceeded to
correct for cluster contamination in our comparison with {\it free} neutron 
data. We subtracted 
contributions to the cross section from composite fragments  by rejecting 
 neutrons  when the distance between the neutron and any other nucleon from
the same BUU ensemble\cite{BUU} 
 is less than a critical distance $d$\cite{aicber85}, which we find to be 
2.5 fm. 
By restricting the analysis to {\it free} neutrons, the BUU calculations 
 of both the double-  and triple-differential 
cross sections agree generally with the data and are insensitive to K;
 the comparison is shown in Fig. \ref{tri24} for the triple-differential 
cross section at 24$^{\circ}$, and in Fig. \ref{doub}  for the 
double-differential cross section. The double-differential cross 
sections are significantly higher (lower) than the data   with $d ~=~ 2.0 (3.0)~fm$.

 The r($\theta$) vs $\theta$ curves (for $ 0.7 < \alpha \leq 1.2$)  in
Fig. \ref{Buupdn} are BUU results for {\it free} neutrons. 
Comparison with the data reveals that r($\theta$) for  {\it free} neutrons is 
also insensitive to K. This result indicates that the azimuthal 
anisotropy  of {\it free} neutrons is not a sensitive probe
of the EOS, contrary to what was hoped previously\cite{Welke88};
however, the fact that full one--body calculations of r($\theta$) exhibit 
considerable structure and sensitivity to K\cite{RM93} points to composites as
the carriers of the information. An important consequence
of our analysis is that the azimuthal anisotropy ratio as a technique to 
probe the EOS is not invalidated by the data on {\it free} neutrons; 
instead, it is necessary to concentrate on the application of this technique 
to composites.  This assertion is presently under close
quantitative scrutiny and preliminary results are 
encouraging\cite{dgp93}. Because of large statistical uncertainties in 
 both the data and the model calculations  in regions of phase space 
 where composite formation is small, K is still constrained poorly 
 in these regions.

The unique features of this experiment include (a) 
the first measurement of neutron flow; (b) simultaneous determination of 
 absolute triple- and double-differential cross sections; and 
(c) a successful demonstration of a new approach to `` 4 $\pi$'' physics 
 at intermediate energies with separation and economical optimization 
 of two previously combined detector functions (viz., measurement of 
$\Phi_R$ and sampling of fragment momenta). 
We simplified the task of filtering models to simulate experimental 
acceptances (with a relatively low cost detector) and measured  
  absolute triple-differential cross sections over a wide energy region.

This work was supported in part by the National Science Foundation under Grant
Nos. PHY-91-07064, PHY-88-02392, and PHY-86-11210, 
the U.S. Department of Energy under Grant No. DE-FG89ER40531 and
DE-AC03-76SF00098, the Natural Sciences and Engineering Research
Council of Canada and the FCAR Nouveaux Chercheurs Fund of the
Qu\'ebec Government. We acknowledge computer time from the Minnesota
Supercomputer Institute and the McGill University Centre for the Physics of
Materials. C.G. and K.H. are happy to acknowledge useful conversations with P.
Danielewicz, S. Das Gupta and V.R. Pandharipande. R. M. thanks M. Gyulassy 
 for suggesting this experiment.

\figure { Triple-differential cross sections for neutrons emitted at 
 a polar angle of  24 degrees with rapidities ($ 0.7 < \alpha \equiv 
(Y/Y_p)_{CM}\leq 1.2$ ) from {\it semi central} Nb-Nb collisions 
 at 400 AMeV with a reaction plane dispersion of 40 degrees.
The ratio of the maximum impact parameter to the nuclear radius  
$b_0/2 R = 0.47$. The broken lines represent results of BUU calculations 
for {\it free} neutrons  with three values of the  incompressibility modulus 
K; the statistical uncertainties are typically $\sim$  10\% 
 for each $\Phi$-bin of 18$^{\circ}$ at polar angles from 
7$^{\circ}$-15$^{\circ}$; increase to $\sim$ 14\% for for $\phi - \phi_R$ $>$ 
90$^{\circ}$  at $\theta$ = 18$^{\circ}$-24$^{\circ}$, and 
become progressively worse as $\theta$ increases from 27$^{\circ}$ to 
36$^{\circ}$.
The solid line represents a least-squares fit to the data.\label {tri24} }
\figure { The polar-angle dependence  of the double-differential cross 
sections for {\it free} neutrons emitted with rapidities 
($ 0.7 < \alpha \equiv (Y/Y_p)_{CM}\leq 1.2$ ) from {\it semi central} 
Nb-Nb collisions at 400 AMeV.
 The circles are the data. The open symbols are BUU calculations.
\label {doub} }
\figure {The polar-angle dependence of the maximum azimuthal anisotropy
ratio $r(\theta)$ for neutrons emitted with rapidities ($ 0.7 <
\alpha \equiv (Y/Y_p)_{CM} \leq 1.2 $) from multiplicity-selected  Nb-Nb 
collisions at 400 AMeV with
impact parameters $ 0 < b(fm) \leq 4.8$.  The circles represent ratios
determined from experiment and corrected to zero dispersion. The ratio of the
 maximum impact parameter to the nuclear radius  $b_0/2 R = 0.47$.
The lines represent results of BUU calculations for {\it free} neutrons 
($d$ ~=~ 2.5~ fm) for four values of the incompressibility modulus K. 
\label {Buupdn} }

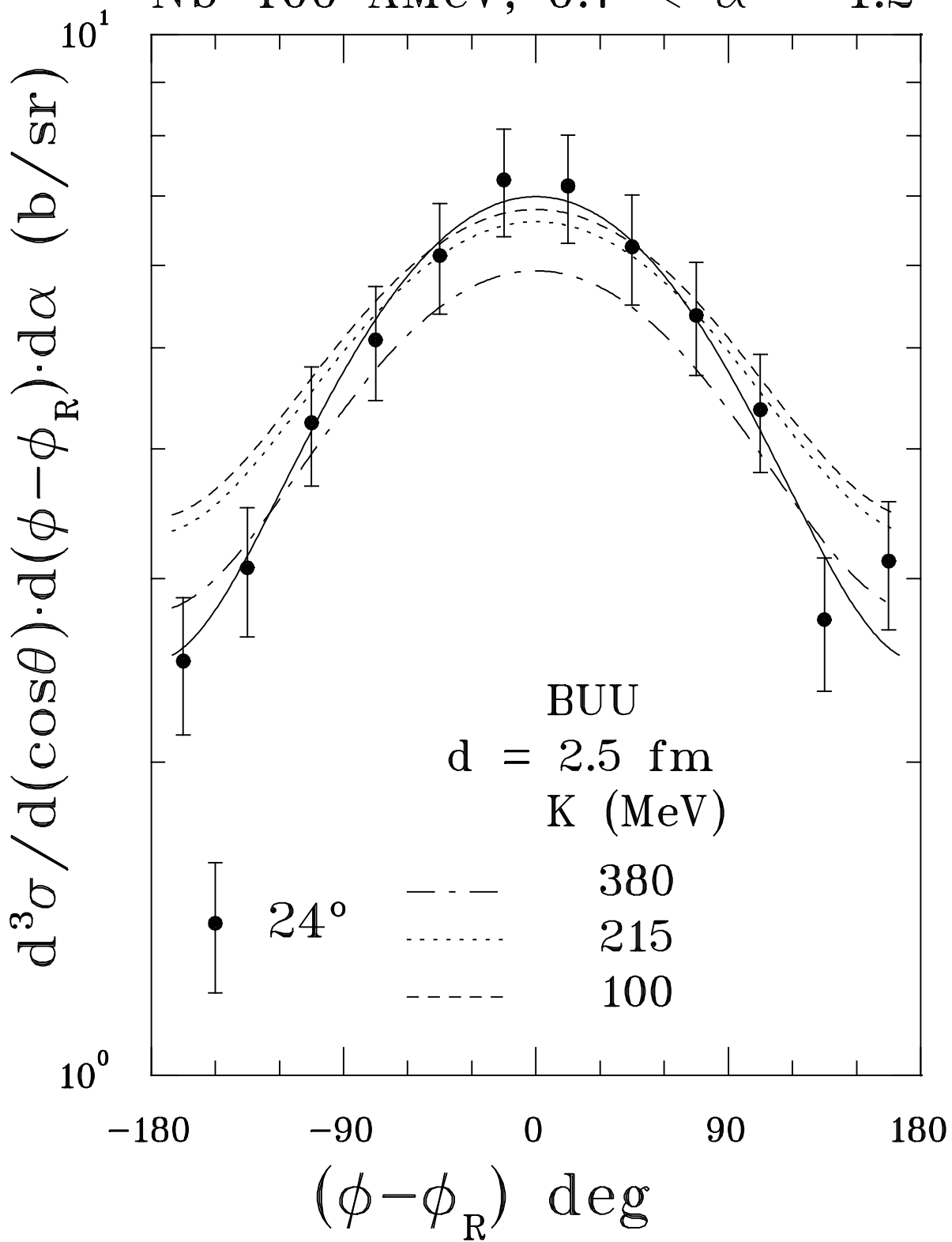
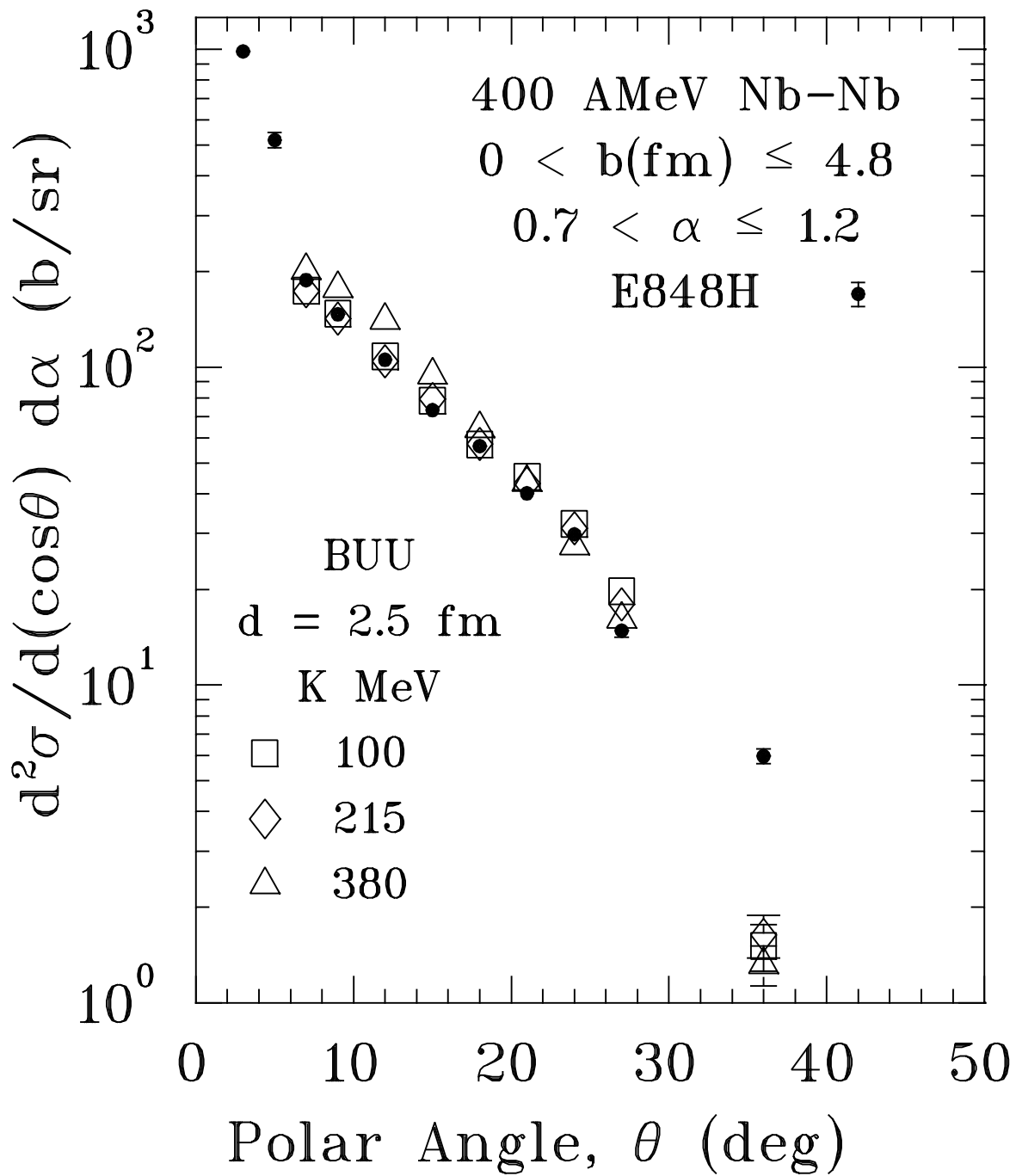
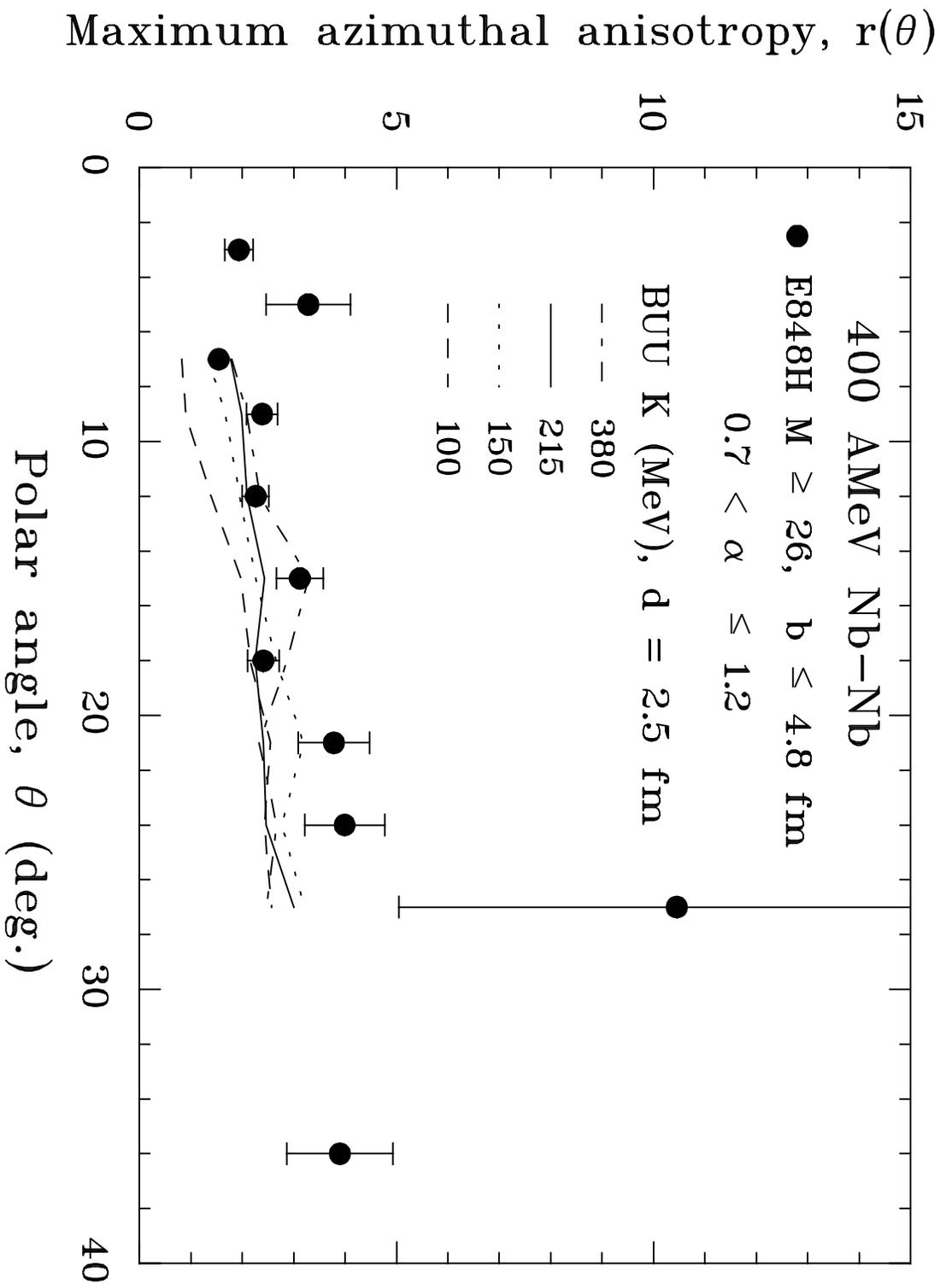
\end{document}